\documentclass[twocolumn,aps,prl,epsfig,showpacs,amsmath,amssymb,superscriptaddress]{revtex4}

\usepackage{graphicx}

\usepackage{dcolumn}
\usepackage{bm}

\begin{document}


\title{Orientational tuning of the 2D-superconductivity in LaAlO$_{3}$/SrTiO$_{3}$ interfaces}

\author{G. Herranz} \email{gherranz@icmab.es}
\affiliation{Institut de Ci\`{e}ncia de Materials de Barcelona (ICMAB-CSIC), Campus de la UAB, Bellaterra 08193, Catalonia,
Spain}

\author{N. Bergeal}
\affiliation{LPEM-UMR8213/CNRS-ESPCI ParisTech-UPMC, 10 rue Vauquelin, 75005 Paris, France}

\author{J. Lesueur}
\affiliation{LPEM-UMR8213/CNRS-ESPCI ParisTech-UPMC, 10 rue Vauquelin, 75005 Paris, France}

\author{J. G\'{a}zquez}
\affiliation{Institut de Ci\`{e}ncia de Materials de Barcelona (ICMAB-CSIC), Campus de la UAB, Bellaterra 08193, Catalonia,
Spain}

\author{M. Scigaj}
\affiliation{Institut de Ci\`{e}ncia de Materials de Barcelona (ICMAB-CSIC), Campus de la UAB, Bellaterra 08193, Catalonia,
Spain}

\author{N. Dix}
\affiliation{Institut de Ci\`{e}ncia de Materials de Barcelona (ICMAB-CSIC), Campus de la UAB, Bellaterra 08193, Catalonia,
Spain}

\author{F. S\'{a}nchez}
\affiliation{Institut de Ci\`{e}ncia de Materials de Barcelona (ICMAB-CSIC), Campus de la UAB, Bellaterra 08193, Catalonia,
Spain}

\author{J. Fontcuberta}
\affiliation{Institut de Ci\`{e}ncia de Materials de Barcelona (ICMAB-CSIC), Campus de la UAB, Bellaterra 08193, Catalonia,
Spain}

\date{\today}

\begin{abstract}

The discovery of a two-dimensional (2D) electron gas at the (110)-oriented LaAlO$_{3}$/SrTiO$_{3}$ interface provided us with the opportunity to probe the effect of crystallographic orientation and the ensuing electronic reconstructions on interface properties beyond the conventional (001)-orientation. At temperatures below 200 mK, we have measured 2D superconductivity with a spatial extension significantly larger ($d$ $\approx$ 24 - 30 nm) than previously reported for (001)-oriented LaAlO$_{3}$/SrTiO$_{3}$ interfaces ($d$ $\approx$ 10 nm). The more extended superconductivity brings about the absence of violation of the Pauli paramagnetic limit for the upper critical fields, signaling the distinctive nature of the electronic structure of the (110)-oriented interface with respect to their (001)-counterparts.

\end{abstract}


\maketitle

In the 60s, type II-superconductivity was found in the wide bandgap semiconductor SrTiO$_3$ (STO) \cite{schooley64, schooley65, koonce67} with transition temperatures below $\sim$ 400 mK that varied non-monotonically over a wide range of low carrier concentrations ($\sim$ 10$^{18}$ - 10$^{21}$ cm$^{-3}$) \cite{koonce67}. The superconductivity in STO exhibited some atypical features. For instance, a particularly large Landau-Ginzburg parameter $\kappa \equiv \lambda/\xi > 10^{2}$ \cite{schooley64} was found, in correspondence to the large penetration depth $\lambda$ $>$ 10 $\mu$m and correlation length $\xi$ $\approx$ 100 nm \cite{koonce67}. On the other hand, the lower critical magnetic fields were fairly small (typically $H_{c1}$ $<$ 5 Oe), whereas the upper critical fields were orders of magnitude larger ($H_{c2}$    $\sim$ 500 Oe) \cite{ambler66}. Recent advances in the field -either via surface electrostatic doping \cite{lee11, ueno11} or growth of STO heterostructures \cite{kozuka09}- have allowed bringing this superconductive state to the two-dimensional limit, where the spatial confinement along one of the dimensions is shorter than the correlation length. Particularly, 2D electron systems at polar (001)-oriented interfaces have been found to be superconductive \cite{reyren07, biscaras10}, with a 2D superconductivity largely modulated by electric gates \cite{caviglia08, biscaras12}. This transition to 2D has led to phenomena not observed in the 3D regime, such as enhanced magnetic-driven superconductivity \cite{gardner11}, violation of the paramagnetic Pauli limit for the upper critical fields \cite{kim12}, quantum phase transitions \cite{schneider09} or multiple quantum criticality \cite{biscaras13}.

\begin{figure}[!h]
\includegraphics [keepaspectratio=true,width=1\columnwidth] {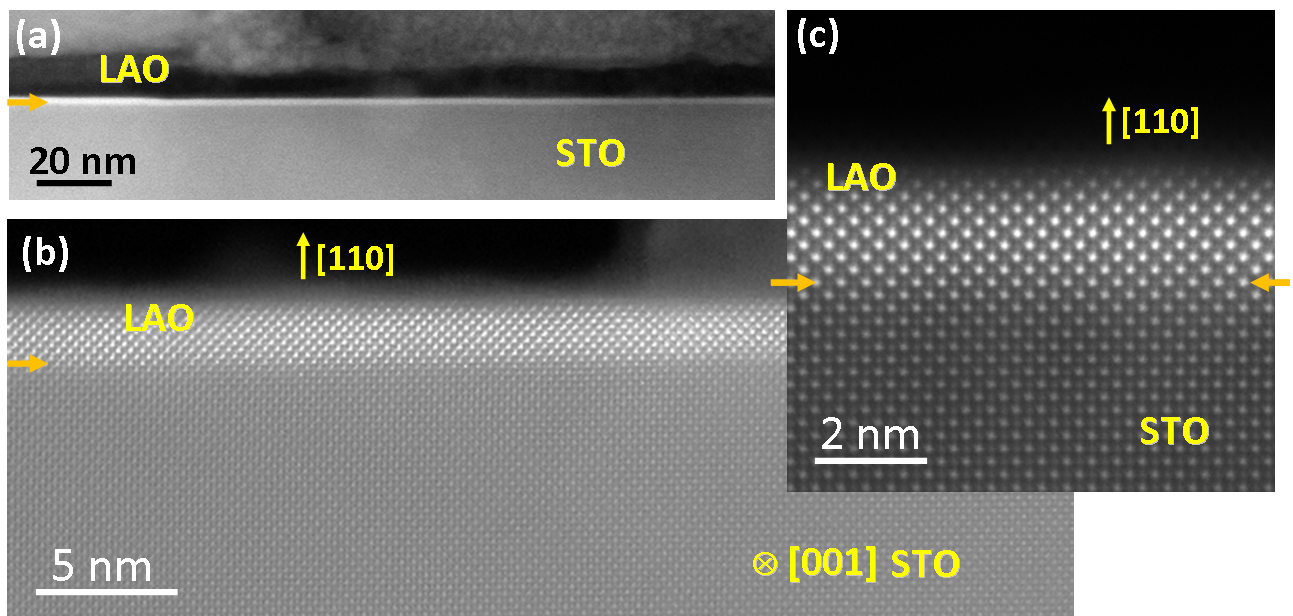} \caption{Cross-sectional HAADF images of the (110)-LAO/STO interface along the [001]-zone axis. (a) Low magnification STEM image showing a continuous film over long lateral lengths. The brighter band, marked by the arrow, corresponds to the nominal 8 MLs thick LAO layer. (b) and (c) high resolution STEM images showing a fully epitaxial film and a coherent interface. The arrows in the images indicate the film-substrate interface. Note the absence of interface faceting. \label{Fig1}}
\end{figure}

Inspired by the possibilities of emerging physics at interfaces beyond (001) \cite{xiao11, ruegg12}, we have recently investigated the transport properties of (110)- as well as (111)-oriented LAO/STO interfaces and found that they sustain high-mobility conduction exhibiting transport parameters in the same range as in (001)-interfaces \cite{herranz12}. An interesting question arises whether 2D superconductivity can also develop along LAO/STO interfaces other than (001)-oriented, and the properties of the distinctive subjacent electronic subband structure. This would allow an unparalleled opportunity to scrutinize the effects of orbital and electronic reconstructions driven by the varying crystallographic orientations on the superconductive electron liquid. In other superconductors -e.g., YBCO, MgB$_{2}$- although the anisotropic crystal structure gives rise to an anisotropy in their electronic properties, the conducting electrons are essentially localized in the CuO$_{2}$ (YBCO) \cite{pickett89} or in the Mg/B (MgB$_{2}$) \cite{buzea01} planes, and thus there is no effective way to tune the electronic structure by changing the crystallographic orientation. The situation is different, however, in STO, which is virtually cubic -at temperatures T $<$ 105 K, STO is tetragonal with a lattice parameter ratio very close to unity $c/a$ $\approx$ 1.0008-, and 2D-like transport might be, in principle, induced either in (001) or (110) planes.

In this Letter we show that indeed (110)-LAO/STO interfaces transit to a 2D superconductive state, as confirmed by the analysis of the electronic transport under magnetic fields using the Ginzburg-Landau theory \cite{tinkham96}. This has given us a unique chance to explore the intrinsic distinctive properties of the electron liquid at the (110)-interface, with respect to those of the (001)-interfaces. As a result, we have observed that the superconductive thicknesses for the (110)-interfaces is significantly larger ($d$ $\approx$ 24 - 30 nm) than those found in their (001)-interfaces. As a further cross-check of this wider extension we analyzed the Pauli paramagnetic limit of the superconductivity, which estimates the maximum upper critical fields on the basis of field-induced orbital pair-breaking \cite{chandrasekhar62, clogston62}. Indeed, for extremely confined STO superconductivity ($t$ $\leq$ 10 nm) the upper critical fields are found to exceed largely this limit, a phenomenon suggested being associated to the strong surface spin-orbit mechanism \cite{kim12}. Here we show, in contrast, that the upper critical fields of the (110)-oriented LAO/STO interfaces are well below the paramagnetic Pauli limit, in full agreement with the significantly wider extension of the electronic liquid with respect to the (001)-interfaces. These experimental facts are then a consequence of the idiosyncratic electron subband hierarchy of the electron liquids confined along the [110] direction.

The samples were obtained by pulsed laser deposition, in which LAO thin films with thickness $t$ = 7 - 14 monolayers (MLs) were grown on thermally treated (110)-oriented STO substrates, see details elsewhere \cite{herranz12, bachelet07}. The interfaces with $t$ $<$ 7 MLs were found to be insulating or poorly conductive, while interfaces with $t$ $\geq$  7 MLs were metallic. The sheet carrier density at 5 K was $n_{sheet}$ $\approx$  1.1 - 10$\times$10$^{13}$ cm$^{-2}$, whereas the electronic mobility was     $\mu$ $\approx$ 200 - 2500 cm$^{2}$/Vs \cite{herranz12}. In the transport experiments the current was injected along the in-plane [001] direction.

Atomic-scale structural characterization of the interfaces - with particular emphasis on the eventual formation of $ \{100\}$ microfacets- is a crucial aspect \cite{bando95, bottin05, annadi13}. For that reason, we carried out high angle annular dark field (HAADF) imaging in a FEI Titan (60-300 kV) scanning transmission electron microscopy (STEM), equipped with a probe-aberration corrector, a monochromator and an XFEG electron gun. Figure \ref{Fig1}a-c shows cross-sectional HAADF-STEM images of the (110)-oriented LAO/STO interface along the [001]-zone axis. In HAADF imaging, the scattering intensity approximately scales with $Z^{2}$, where $Z$ is the atomic number, whereby the brighter atomic columns correspond to the heavier elements. In these images, an epitaxial LAO layer with thickness $t$ $\sim$ 8 MLs is clearly visible, being continuous over long lateral lengths. Besides, along this zone axis the (110) ionic stacking across the interface can be readily appreciated, see Figs. \ref{Fig1}b and \ref{Fig1}c, demonstrating that (110)-oriented LAO/STO interface is atomically flat, ruling out the presence of any reconstructed surfaces with local $ \{100\}$ microfacets.

\begin{figure}[!h]
\includegraphics [keepaspectratio=false,width=0.85\columnwidth] {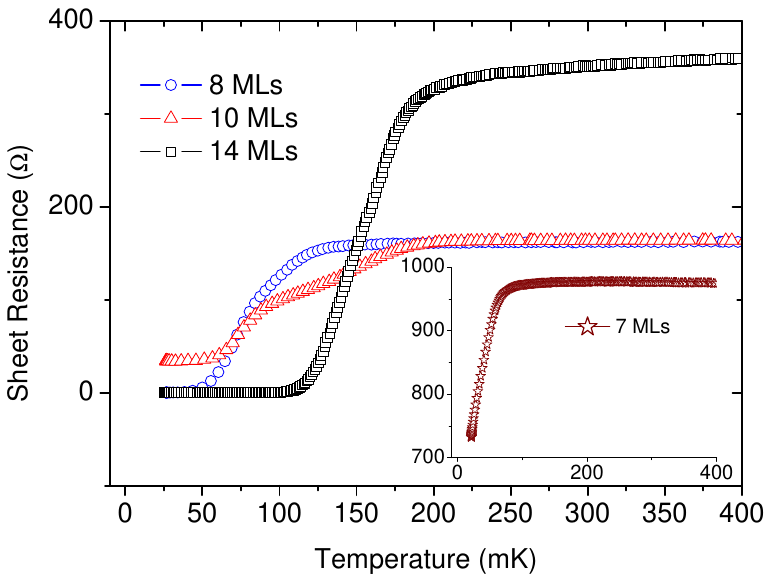} \caption{Sheet resistance versus temperature for (110)-oriented LAO/STO interfaces with LAO thickness $t$ = 8, 10 and 14 MLs. The inset displays the sheet resistance corresponding to $t$ = 7 MLs. \label{Fig2}}
\end{figure}

To investigate the superconductive state we measured the sheet resistance of the heterostructures down to 20 mK. As seen in Figure \ref{Fig2}, for all the analyzed samples the resistance dropped drastically below a critical temperature $T_{c}^{onset}$. The onset of the transition is defined here as the temperature where the resistance falls to 90\% of its normal state value (measured at T = 400 mK). We observed that $T_{c}^{onset}$ was strongly dependent on the sample, being the highest for $t$ = 14 MLs ($T_{c}^{onset}$ $\approx$ 195 mK), whereas lower values were observed for $t$ = 10 MLs ($T_{c}^{onset}$ $\approx$ 166.5 mK), $t$ = 8 MLs ($T_{c}^{onset}$ $\approx$ 115 mK) and $t$ = 7 MLs ($T_{c}^{onset}$ $\approx$ 45 mK). Samples with $t$ = 14 MLs and $t$ = 8 MLs exhibited the narrowest transition widths $\Delta T_{C}$ $\simeq$ 40 mK (defined between 20\% and 80\% of the normal state resistance), comparable to the $\Delta T_{C}$ observed in superconductive (001)-LAO/STO as well as (001)-LaTiO$_{3}$/STO interfaces \cite{reyren07, biscaras10, biscaras12}. The other samples exhibited significantly broader transitions, indicating more inhomogeneous electronic states. This observation is more perceptible in sample $t$ = 10 MLs that exhibits a complex transition to superconductivity with two kinks (Figure \ref{Fig2}).

\begin{figure}[!h]
\includegraphics [keepaspectratio=false,width=1\columnwidth] {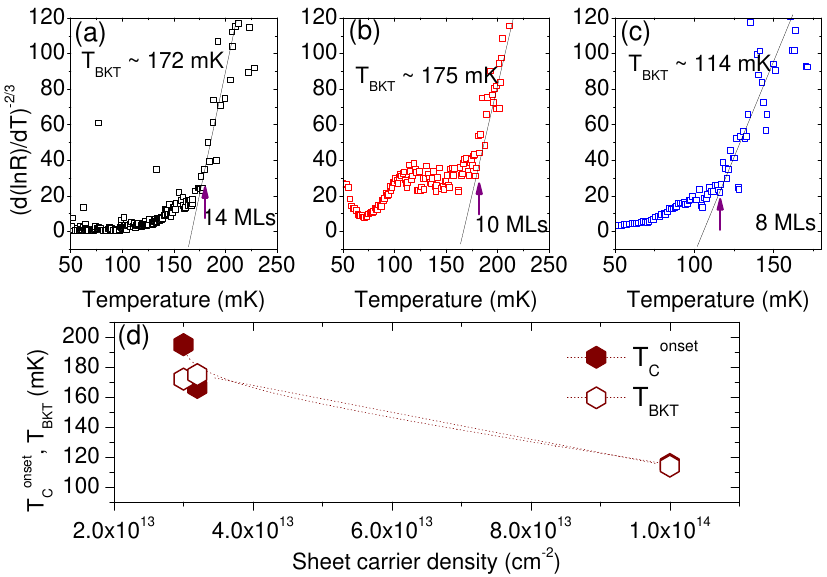} \caption{Temperature dependence of [$\partial$ln$R$/$\partial$$T$]$^{2/3}$ for samples with thickness (a) $t$ = 14 MLs, (b) $t$ = 10 MLs and (c) $t$ = 8 MLs. Arrows signal the temperature $T_{BKT}$ inferred from the fittings. The superconducting transition temperatures $T_{c}^{onset}$ and $T_{BKT}$ are plotted as a function of the sheet carrier density in (d). \label{Fig3}}
\end{figure}

In the 2D limit, it is well established that superconductivity is suppressed by vortex-antivortex pair unbinding by thermal fluctuations, so that the transition to the 2D superconductive state belongs to the Berezinskii-Kosterlitz-Thouless (BKT) universality class \cite{kosterlitz73, beasley79}. In this scenario, above the transition temperature $T_{BKT}$ the free motion of vortices triggers the transition from the 2D superconductive towards the normal state. Within a narrow region around $T_{BKT}$ the BKT theory predicts specific laws for the temperature dependence of the resistance $R$, giving a quantifiable criterion to assess the dimensional character of the superconductive state. In particular, it is expected that [$\partial$ln$R$/$\partial$$T$]$^{2/3}$ should scale linearly with ($T$ - $T_{BKT}$) for a certain region around $T_{BKT}$ \cite{she12}. As seen in Figures \ref{Fig3}a-c, similarly to (001)-LAO/STO interfaces, the [$\partial$ln$R$/$\partial$$T$]$^{2/3}$ data fall approximately on a straight line above a certain temperature that -following the same protocol as described in the analogous analysis of (001)-interfaces \cite{reyren07, caviglia08, schneider09}- we identify as the $T_{BKT}$ transition temperature. Therefore, the temperature $T_{BKT}$ is at the boundary between the two regimes (superconductive/normal) giving rise to a prominent kink in the [$\partial$ln$R$/$\partial$$T$]$^{2/3}$ plot. We note that the values of $T_{BKT}$ thus obtained are in the same range as those found in (001)-LAO/STO \cite{reyren07}. We find that the $T_{BKT}$ values correlate with $T_{c}^{onset}$ (Figure \ref{Fig3}d), both being strongly dependent on the sheet carrier density $n_{sheet}$ in the normal state, a feature that is reminiscent of the strong dependence of the bulk superconductive transition on carrier density \cite{koonce67}.

\begin{figure}[!h]
\includegraphics [keepaspectratio=false,width=1\columnwidth] {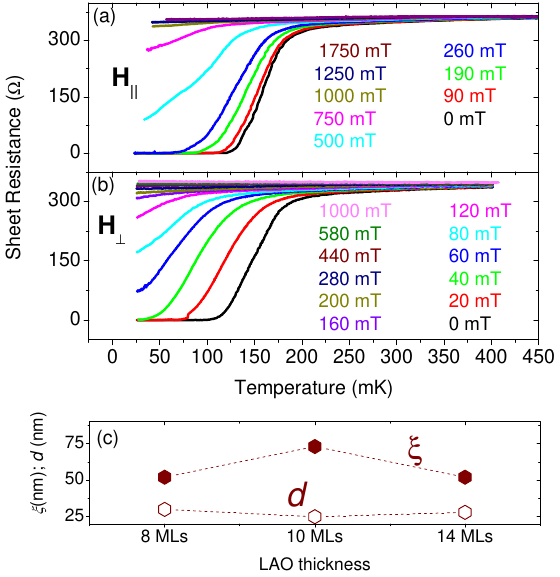} \caption{Sheet resistance of the (110)-interfaces measured for $t$ = 14 MLs as a function of temperatures for magnetic field applied (a) parallel ($H_{\|}$ ) and (b) perpendicular ($H_{\bot}$) to the interface. The coherence length ($\xi$) and superconductive thickness ($d$) are plotted in (c) as a function of the LAO overlayer thickness. \label{Fig4}}
\end{figure}

In order to assess further the 2D character of the superconductive state, we measured the resistance as a function of temperature for different values of magnetic fields applied both parallel and perpendicular to the interface. As illustrated in Figure \ref{Fig4} (sample with $t$ = 14 MLs), a high enough magnetic field progressively induces the transition from the superconductive to the normal state. However, a first inspection of this Figure immediately reveals that the field required for this transition is very different depending on the orientation. The resistance goes to the normal state value above 120 mT when the field is applied normal to the interface (Figure \ref{Fig4}b), while significantly higher fields above 750 mT are required in-plane (Figure \ref{Fig4}a). This strong anisotropy is indeed a hallmark of low-dimensional superconductivity, since in the 2D limit superconductivity cannot be suppressed by vortex entry for in-plane fields \cite{kozuka09}.

\begin{figure}[!h]
\includegraphics [keepaspectratio=false,width=0.8 \columnwidth] {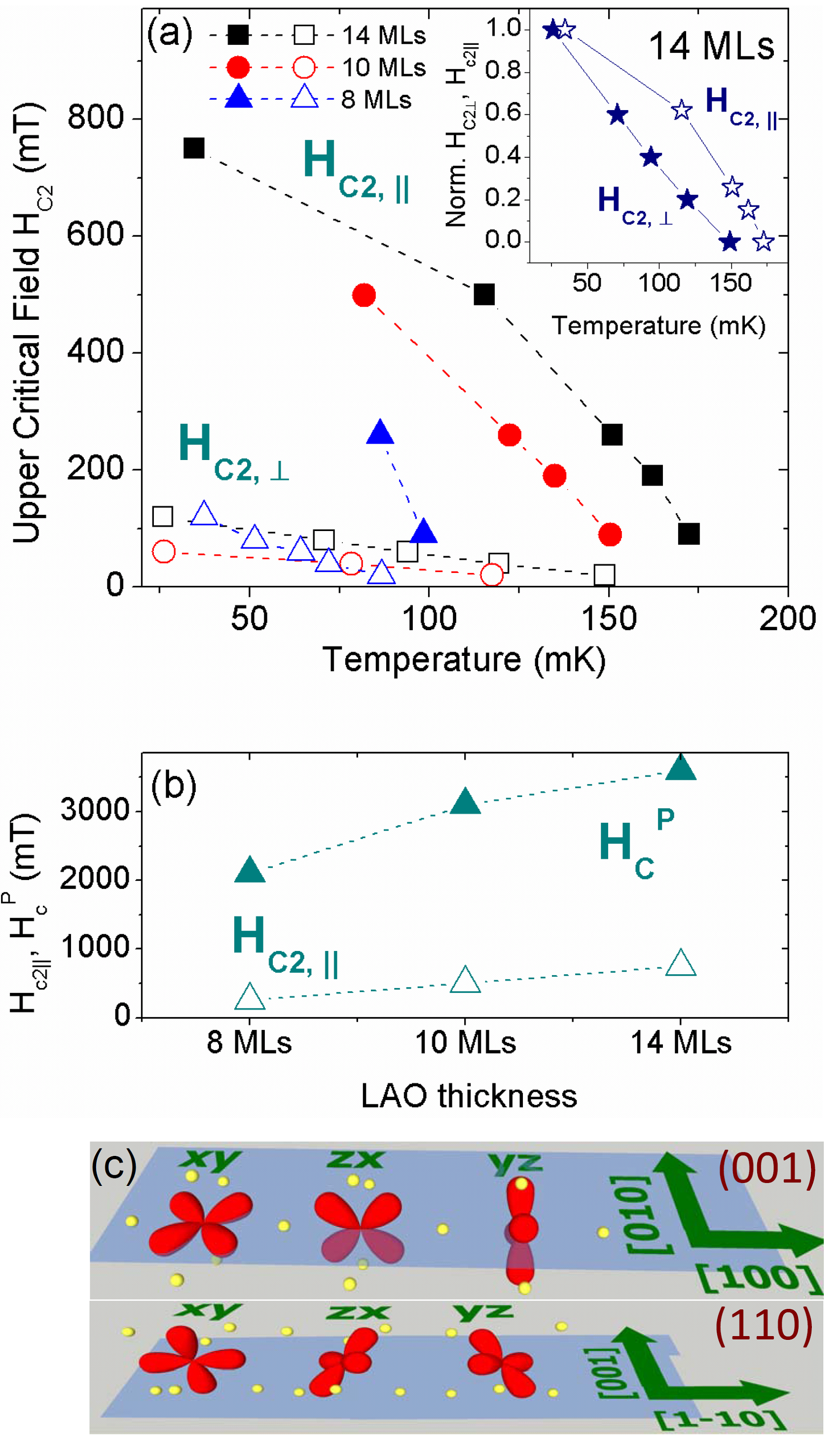} \caption{(a) Upper parallel ($H_{c2,\|}$) and perpendicular ($H_{c2,\bot}$) critical fields as a function of the temperature. Inset: Temperature dependence of the normalized $H_{c2,\|}$ and $H_{c2,\bot}$ of sample $t$ = 14 MLs. (b) Upper parallel critical field $H_{c2,\|}$ and Pauli-limited critical field $H_{c}^{P}$ as a function of the LAO overlayer thickness (c) Sketch of the $t_{2g}$ orbitals in (001)- and (110)-oriented interfaces. The yellow spheres indicate some of the oxygen lattice positions. \label{Fig5}}
\end{figure}

A more detailed analysis, following the Landau-Ginsburg formalism, can indeed give a quantitative estimation of both the superconductive layer thickness $d$ and the in-plane superconductive coherence length $\xi$ \cite{tinkham96}. For that purpose, the upper normal $\mu_{0}$$H_{c2,\bot}$ and in-plane $\mu_{0}$$H_{c2,\|}$ critical fields have to be determined by defining quantitative criteria for the field-induced transitions. Thus, a drop resistance of 90\% from the normal resistance state at $T$ = 400 mK was established to ascertain the evolution of $T_{c}^{onset}$ with the applied fields. With this criterion, we observed that for $t$ = 14 MLs the upper normal critical field extrapolated to $T$ = 0 K was $\mu_{0}$$H_{c2,\bot}$ $\approx$ 160 mT, leading to an in-plane coherence length $\xi = \sqrt{\frac{\Theta_{0}}{2\pi\mu_{0}H_{c2,\bot}}}$ = 44 nm ($\Theta_{0}$ is the flux quantum) \cite{tinkham96}. Measurements performed in a parallel geometry (Figure \ref{Fig4}a) gave instead a critical field $\mu_{0}$$H_{c2,\|}$ $\approx$ 1000 mT, so that we could estimate the superconductive thickness $d = \frac{\Theta_{0}\sqrt{3}}{\pi\mu_{0}H_{c2,\|}\xi}$ = 24 nm for $t$ = 14 MLs. Similarly, the in-plane coherence length and electron liquid thicknesses were obtained for the rest of samples applying the same protocol, obtaining $\xi$ $\approx$ 40 - 75 nm and $d$ $\approx$ 24 - 30 nm (these values are plotted in Figure \ref{Fig4}c for all samples). An alternative guess of both $\xi$ and $d$ can be obtained from the analysis of the temperature dependence of the critical fields $H_{c2,\|}$ and $H_{c2,\bot}$ (Figure \ref{Fig5}a). For a 2D superconductor, the upper critical field in the perpendicular geometry is given by $\mu_{0}$$H_{c2,\bot} = \frac{\Theta_{0}}{2\pi\xi^{2}}(1-T/T_{c}^{onset})$, whereas in the parallel geometry we have $\mu_{0}$$H_{c2,\|} = \frac{\Theta_{0}\sqrt{12}}{2\pi\xi d}(1-T/T_{c}^{onset})^{1/2}$ \cite{kozuka09}. These dependences are clearly observed for $t$ = 14 MLs. By fitting the experimental curves, and recalling that $T_{c}^{onset}$ $\approx$ 195 mK for $t$ = 14 MLs, we obtained $\xi$ $\approx$ 45 nm and $d$ $\approx$ 26 nm, in perfect agreement with the previous analysis.

How it compares the superconductive state at the (110)-LAO/STO interfaces with respect to that at (001)-LAO/STO? We note that whereas the coherence lengths for (001)-LAO/STO are found to be $\xi$ $\approx$ 70 - 100 nm \cite{reyren07}, in (110)-LAO/STO we find significantly smaller values, $\xi$ $\approx$ 40 - 75 nm. On the other hand, the superconductive layer is found to be appreciably larger in (110)-LAO/STO ($d$ $\approx$  24 - 30 nm) than in (001)-LAO/STO ($d$ $\approx$ 10 nm \cite{reyren07, reyren09}). However, although strongly indicating the more extended nature of the electron liquid at (110)-interfaces, it is generally accepted that the Landau-Ginsburg analysis gives an upper bound for the superconductive thickness \cite{shalom10}. We require, therefore, additional compelling evidence to make the case for the (110)-oriented interfaces.

An additional cross-check of the wider spatial extension of the (110)-2D state is provided by the analysis of the Pauli paramagnetic limit of the upper critical fields. For high enough magnetic fields, the paramagnetic susceptibility induces a parallel alignment of the Cooper pair spins that eventually breaks them apart, giving a higher limit for the upper critical fields. This limit can be assessed as $\mu_{0}$$H_{c}^{P} = 1.76k_{B}T_{c}^{onset}/\sqrt{2}\mu_{B}$  (Eq. 1) \cite{kim12, chandrasekhar62}, where $k_{B}$ is the Boltzmann's constant and $\mu_{B}$ is the Bohr magneton (assuming a $g$ factor of 2). Although this limit is generally fulfilled, it is violated in some exceptional cases. One example is the case of ultrathin STO 2D superconducting layers with thickness below $d$ $\approx$ 8.8 nm, for which the values of $\mu_{0}$$H_{c2,\|}$ were found to exceed largely the Pauli limit, signaling the relevance of intrinsic spin-orbit coupling effects for extremely confined superconductivity in STO \cite{kim12}. Indeed, the paramagnetic limit is also violated in (001)-oriented LAO/STO interfaces \cite{shalom10}, where the superconductive thickness has been consistently determined to be $d$ $\approx$ 10 nm \cite{reyren07, reyren09}. The large in-plane critical fields up to $\mu_{0}$$H_{c2,\|}$ $\approx$      2 T found in (001)-LAO/STO interfaces \cite{reyren09} -much larger than those found here for the (110)-interfaces- are coherent with this interpretation. In contrast, the significantly spatially wider superconductivity at the (110)-LAO/STO interfaces should avoid the violation of the paramagnetic Pauli limit. Indeed, according to Kim et al., the upper critical in-plane fields $\mu_{0}$$H_{c2,\|}$ in superconductive STO thin films do not violate the Pauli limit when the thickness is above $d$ $\approx$  20 nm (see Figure 3 in Ref. \cite{kim12}). In agreement with these expectations, we have observed that the measured in-plane critical fields were $\mu_{0}$$H_{c2,\|}$ $<$ 750 mT for all samples, well below the Pauli limited critical fields $\mu_{0}$$H_{c}^{P}$ $\approx$ 2000 - 3500 mT calculated according to Eq. 1 (see Figure \ref{Fig5}b).

We suggest that the ultimate reason for the increased spatial extension is the subband reconstruction driven by the electron confinement along the [110] direction. For the conventional (001)-interfaces the subband structure consists of a couple of degenerate, strongly dispersive bands with light mass electrons and a weakly dispersive (heavy-mass) band \cite{santander11}, with the lowest band having a $d_{xy}$ character at the $\Gamma$-point and $d_{xz}/d_{xy}$ bands lying higher in energy. \textit{Ab-initio } calculations show that for that particular orientation the $d_{xz}/d_{xy}$ bands extend significantly farther away from the interface than the $d_{xy}$ states, which remain more strongly bound \cite{delugas11, popovic08}. So far, the electronic and orbital reconstructions associated to the confinement of the electron liquid along the [110] remain unexplored. However, the clearly distinctive spatial distribution of $t_{2g}$ states with respect to the interface (see the sketch in Figure \ref{Fig5}c) anticipates a modified energy subband hierarchy of $d_{xy}$ and $d_{xz}/d_{xy}$ states and/or a renormalization of the associated effective band masses. We propose that such interface-reorientation driven electronic reconstruction is behind the larger spatial extension of the 2D electron liquid at the (110)-interface. At the same time, our results open up fascinating novel questions, e.g., how are the coexistence of magnetism and superconductivity or the Rashba spin-orbit coupling affected by the reorganized electronic hierarchy and larger extension of the electron system in the (110)-interfaces.

\begin{acknowledgments}

This work was supported by the Spanish MAT2011-29269-C03 and NANOSELECT CSD2007-00041 projects and the Generalitat de Catalunya (2009 SGR 00376 project), and by the R\'{e}gion \^{I}le-de-France in the framework of CNano IdF program. The HRTEM microscopy work was conducted in "Laboratorio de Microscop\'{i}as Avanzadas" at the Instituto de Nanociencia de Arag\'{o}n-Universidad de Zaragoza. Authors acknowledge the LMA-INA for offering access to their instruments and expertise. JG is grateful to Dr. A. Ibarra for the help at LMA-INA. Also fruitful discussions with Mar\'{i}a Varela are acknowledged.

\end{acknowledgments}

\bibliography{LAOSTO}

\end{document}